\documentclass[aps,preprint,eqsecnum,prd]{revtex4}
\usepackage{epsfig}

\newcommand{\beq}{\begin{equation}}
\newcommand{\eeq}{\end{equation}}

\newcommand{\vev}{{\langle\, \phi\, \rangle}}
\newcommand{\vevphi}{{\langle\ \phi\ (\varphi)\ \rangle}}
\newcommand{\vevphial}{{\langle\ \phi\ (\varphi, \alpha)\ \rangle}}

\newcommand{\remove}[1]{}
\newcommand {\comment}[1]{}             
\begin{document}
%{\tighten
%\preprint{\vbox{\hbox{EUPC/00--01}}}

%\baselineskip 16pt
%\renewcommand\baselinestretch{1.2}

\title{Topologically Alice Strings and Monopoles}

\author{Katherine M.~Benson}
\email{benson@physics.emory.edu}
\affiliation{
Department of Physics,
Emory University,  400 Dowman Drive, Suite N201,Atlanta, GA\ 30322}
\author{Tom Imbo}\email{imbo@uic.edu}
\affiliation{Department of Physics,
University of Illinois at Chicago,
845 W. Taylor St,  m/c 273,
Chicago, IL 60607-7059}

%\bigskip
\date{April 9, 2004}

\begin{abstract}
Symmetry breaking can produce ``Alice'' strings, which alter scattered
charges and carry monopole number and charge when twisted into loops.
Alice behavior arises algebraically, when strings obstruct unbroken
symmetries --- a fragile criterion. We give a topological criterion,
compelling Alice behavior or deforming it away. Our criterion, that
$\pi_o(H)$ acts nontrivially on $\pi_1(H)$, links topologically Alice
strings to topological monopoles. We twist topologically Alice loops
to form monopoles. We show that Alice strings of condensed matter
systems (nematic liquid crystals, $^3$He-A, and related non-chiral
Bose condensates and amorphous chiral superconductors) are
topologically Alice, and support fundamental monopole charge when
twisted into loops. Thus they might be observed indirectly, not as
strings, but as loop-like point defects. We describe other models,
showing Alice strings failing our topological criterion; and twisted
Alice loops supporting deposited, but not fundamental, monopole
number.

\end{abstract}

\maketitle

%\renewcommand{\baselinestretch}{1.6}

%\pacs{11.27.+d, 11.30.Fs, 11.30.Ly,98.80.Cq}
% end the tighten
%\narrowtext
%\newpage

%\renewcommand{\baselinestretch}{1.6}
\section{Introduction}

Among the defects created when gauged symmetries break down are Alice
strings.\cite{oldAlice,stringzm,newAlice,blp} Alice strings obstruct
the global extension of unbroken symmetries, making them multivalued
when parallel transported around the string. This algebraic
obstruction has two prominent physical consequences. First, it
produces nonconservation of associated charges, when Aharonov-Bohm
scattered around the string. Second, it induces monopoles, as twisted
loops of Alice string.

These Alice features arise due to gauge flux on the string's core.
The gauge flux generates the condensate winding, while acting on
asymptotic particles through the Wilson line $U(\varphi)$. This action
fixes particles' Aharonov-Bohm scattering around the string, to one
changing both charge and monopole number. Loops of string, which
leave charge and monopole number well-defined asymptotically, thus
support deposited unlocalizable charge (``Cheshire charge'') and
deposited monopole number. \cite{oldAlice,stringzm} Alice loops carry
this deposited monopole number by twisting, \cite{blp} as we probe
further below.

Alice strings in condensed matter systems are global, not gauged,
defects. They have no gauge flux to fix their Aharonov-Bohm
scattering, and guarantee altered charge and monopole number upon
string traversal. However, their Aharonov-Bohm scattering was
considered in \cite{ABrefs, global}; with \cite{global} showing that
global Alice strings generically share all Alice behaviors.  They
alter both charge and monopole number on string traversal, with
twisted loops supporting both Cheshire charge and deposited monopole
number. Furthermore, condensed matter systems offer the most likely
prospects for Alice string observation. \cite{volbook} We show here
that known condensed matter Alice strings form twisted loops with
fundamental monopole charge, suggesting a second avenue for 
potential observation: Alice strings might be observed, not as
strings, but as looplike point defects, when twisted loops comprise
the energetically favored solution of fundamental monopole charge.

The criterion for Alice string formation was first stated
algebraically, in terms of the string's untraced Wilson loop
$U(2\pi)$.  When $U(2\pi)$ fails to commute with an unbroken symmetry
$h$, the symmetry cannot be globally extended; when it fails to
commute with unbroken generator $T_h$, the associated charge is
nonconserved. Thus Alice strings arise when $U(2\pi)$ lies outside the
center of the unbroken symmetry group $H$. As noted in
\cite{stringzm}, this is an inherently nontopological criterion, as
topologically equivalent choices for $U(2\pi)$ can commute with
different subgroups of $H$. Thus
emergence of Alice behavior appeared a dynamical question. Steps
toward topologizing this criterion came in \cite{blp}. They noted that
when {\bf all} topologically equivalent choices for $U(2\pi)$ lie
outside the center of $H$, Alice strings must form. Equivalently,
topologically Alice strings form when the fiber bundle of $H$ parallel
transported around the string is nontrivializable. Both criteria,
while accurate, seem difficult to apply.

We here establish an easily applied topological criterion which states
when Alice strings {\bf must} form.  Consider the symmetry breakdown of Lie group $G \rightarrow H$, taking for $G$ the simply connected cover of the initial Lie symmetry. A
topological string then has homotopy
$\pi_o(H)$; that is, its flux $U(2\pi)$ lies in a disconnected component
of the unbroken symmetry group $H$.  Monopoles have homotopy $\pi_1(H)$, describing loops
$h(\alpha)$ of different winding in $H$. Our criterion labels strings
with flux $U(2\pi)$ ''topologically Alice'' if they alter the topological
charge of monopoles circumnavigating them:
$$h(\alpha) \rightarrow \tilde{h} (\alpha) = U(2\pi)\ h(\alpha)\
U^{-1}(2\pi)\ \not\sim \ \ h(\alpha) \ .$$ This is a topological criterion,  corresponding to a nontrivial action of $\pi_o(H)$ on $\pi_1(H)$, where
$h_o = U(2\pi) \in \pi_o(H)$ alters the topological winding of loop
$h(\alpha) \in\pi_1(H)$: \beq \tilde{h}
(\alpha) = h_o\ h(\alpha)\ h_o^{-1} \ \ \not\sim \ \ h(\alpha)\ \
.\eeq

This criterion captures physical Alice behavior and is easily applied.
It assures the standard Alice constellation of behaviors: multivalued
symmetry, charge-violating Aharonov-Bohm scattering, Cheshire charge
on Alice loops. In topology lies both power and limitation. As
limitation, Alice behavior survives our criterion {\bf only} when
altering generators alters the topology of loops they generate,
eliminating many models with algebraic Alice phenomena. Most simply,
for $\pi_o(H)$ to act nontrivially on $\pi_1(H)$, $\pi_1(H)$ itself
{\bf must be nontrivial}.  Thus, only a theory with topological
strings and monopoles --- indeed, with monopoles topologically
distinct from antimonopoles --- can Alice strings meet our
criterion. Strings meeting our criterion, however, have Alice behavior
which is topologically ordained; while those failing can be
destabilized, continuously deformed to remove all Alice
behavior. Their Alice behavior survives  {\bf only}
if dynamically stabilized; that is, if energetically favored over
non-Alice strings of the same winding.

Our topological criterion for Alice strings ensures that twisted loops
carry monopole charge, as we see by explicit construction in section
\ref{monopoles}.  We note that topological arguments indicate only 
that {\bf deposited} monopole charge can be carried by a twisted Alice
loop. Monopole charge, however, is typically not deposited in single
fundamental units, leaving open the question of whether twisted Alice
loops can carry {\bf fundamental} monopole charge. Our 
analysis of this model-dependent question gives the heuristic
answer: twisted Alice loops in condensed matter systems typically
support fundamental monopole charge; while twisted Alice loops of
particle physics models often do not.

Specifically, we address the best condensed matter candidates for
Alice strings: the original Alice string of Schwarz, coinciding with
the Alice string of liquid crystals and of non-chiral Bose condensates
\cite{oldAlice, leonvol}; and the Alice string of $^3$He-A
\cite{volmin}, arising anew for unconventional spin-triplet
superconductors \cite{spintrips}. In both models, strings are
topologically Alice, and form twisted Alice loops supporting
fundamental monopole charge, even though monopole charge deposits onto
string loops in even increments.  Thus, for the Alice strings of
interest to condensed matter, even the most fundamental singular point
defect, or monopole, can take the form of a twisted Alice loop.  We
display particle physics-motivated models, however, whose twisted
Alice loops support only deposited monopole charge, and
cannot form fundamental monopoles.

We present our results as follows.  In section \ref{top}, we discuss
our topological criterion for Alice behavior. We show that Alice
strings failing our criterion have topologically unstable Alice
features; that is, their Alice behavior can be deformed away.  In
section \ref{monopoles}, we show, by construction, that twisted Alice
strings carry monopole charge.  We argue topologically, using our
criterion to display a fully twisted Alice loop, carrying monopole number
deposited in the monopole scattering $h(\alpha) \rightarrow
\tilde{h}(\alpha)$. We then illustrate our criterion and Alice loop
twisting to form monopoles, for key models.  First, we treat the
condensed matter systems: in section \ref{schwarz}, the simple Schwarz
Alice string, coinciding with the Alice string of liquid crystals and
some non-chiral Bose condensates; and in section \ref{he3a}, the Alice
string of $^3$He-A and spin-triplet superconductors.  In both cases,
topologically Alice strings twist into loops carrying fundamental
monopole charge; in both cases, this stems from the embedding of monopole
loops in an $SO(3)$ symmetry group, an embedding endemic to condensed
matter systems. This suggests that for all topologically Alice
condensed matter systems, fundamental point defects --- monopoles --- can
take toroidal shape, as twisted Alice loops.

In the particle physics models of section \ref{partmodels}, we
illustrate different outcomes for the topology of Alice strings and
their twisted loops: Alice behavior which is nontopological; and
topologically Alice loops which support deposited, but not
fundamental, monopole charge.  Key points include a focus on how
algebraically Alice candidates may fail our criterion; and how the
half-twisted Alice loops capable of supporting fundamental monopole charge
fail to be single-valued. Thus only fully twisted Alice loops arise,
supporting only the deposited monopole charge dictated by topology.

We conclude in section \ref{conclusions}.

\section{\label{top} A Topological Criterion for Alice Behavior}

We take $G$ to be the simply
connected cover of the initial symmetry --- a connected Lie group ---
and $H\subset G$ its unbroken subgroup.
A topologically stable string has flux
$U(2\pi)$ in a disconnected component of $H$, with
 topology determined by $\pi_o(H)$. Similarly,  the
topology of the monopole is given by $\pi_1(H)$, describing loops
$h(\alpha)$ of different winding in $H$. By taking seriously the
change in  monopole number in circumnavigating the Alice
string, we construct our criterion. Note that, in Aharonov-Bohm
scattering around the string, the monopole $h(\alpha)$ is conjugated
by the string's Wilson loop $U(2\pi)$:
$$h(\alpha) \rightarrow \tilde{h} (\alpha) = U(2\pi)\ h(\alpha)\
U^{-1}(2\pi)\ \ .$$ Monopole number changes if $\tilde{h}(\alpha)$ and
$h(\alpha)$ are topologically distinct loops. We
represent this transformation topologically, as $\pi_o(H)$ acting
naturally on $\pi_1(H)$ by conjugation. Topologically Alice strings form if
that action is nontrivial: that is, if, for $h_o$ a representative
element of $\pi_o(H)$ and $h(\alpha)$ a representative loop in
$\pi_1(H)$,
\begin{equation} \tilde{h} (\alpha) = h_o\ h(\alpha)\ h_o^{-1} \ \ \not\sim \ \
h(\alpha)\ \ .\label{crit}\end{equation}
A string with untraced Wilson loop $U(2\pi) \ \sim \ h_o$
meeting this criterion is topologically guaranteed to change monopole
number; we dub it ``topologically Alice''.

This criterion captures physical Alice behavior, is easily applied,
and is topological. Its result, for any chosen $ h_o$ and $h(\alpha)$,
remains invariant under deformations of either flux $U(2\pi) = h_o$ or
monopole loop $h(\alpha)$. By construction, strings are topologically
Alice if monopoles change topologically in traversing them. Of course,
monopoles $h(\alpha)$ change because of algebraic Alice behavior: loop
$h(\alpha) = e^{i\alpha T_h}$ alters only if its generator $T_h$
alters; that is, if $U(2\pi)$ fails to commute with generator
$T_h$. This algebraic noncommutation creates the standard Alice
constellation of behaviors: multivalued symmetry, charge-violating
Aharonov-Bohm scattering, Cheshire charge on Alice loops. It is
captured by our criterion {\bf only} when altering generators alters
the topology of the loops they generate. This misses some Alice
phenomena --- particularly in models with poorly distinguished loops,
when $\pi_1(H) = 0$ (and all loops are trivial) or $\pi_1(H) = Z_2$
(and all nontrivial loops, including a loop and its inverse, are
identified). Such models possess either no, or only $Z_2$, topological
monopoles.  We claim that Alice behavior in these models is not robust
topologically; that is, continuous deformation of such strings removes
their Alice behavior. In such cases, persistence of Alice behavior can
arise only from dynamical arguments, favoring Alice strings over
non-Alice strings of the same winding. Dynamically stabilized features
remain interesting --- for example, nontopological defects including
embedded, semilocal, or electroweak strings. \cite{semew} However, we
seek here for Alice behavior the more robust motivation of topological
imperative.

Consider an Alice string with Wilson loop $U(2\pi) = h_o$ which 
fails our topological criterion. This occurs if, for a nontrivial loop
$h(\alpha)\ \in\ H$ describing a monopole, the parallel transported
monopole is homotopic to  the original; that
is,
\beq\tilde{h} (\alpha) = h_o\ h(\alpha)\ h_o^{-1} \ \ \sim \ \
h(\alpha)\ \ .\label{nocrit}\eeq
This occurs only if there exists some continuous map $f(x)$ deforming $\tilde{h} (\alpha)$ to $h(\alpha)$; that is
$$f(x) \ :\ \tilde{h} (\alpha) \ \ \rightarrow\ \ \left\{
\ \begin{array}{cc}
\tilde{h} (\alpha) & \mbox{when $x=0$} \\ h(\alpha) &\mbox{when $x=1 $}
\end{array}\ \ . \right.  $$
Note that the map $f(x)$
relates nontrivial loops in $H$, with basepoint $\alpha = 0$ fixed at the
group identity.  We write it  as a continuously varying group element $f(x)$ acting on $\tilde{h} (\alpha)$ by conjugation,
$$f(x) \ :\ \tilde{h} (\alpha) \ \ \rightarrow\ \ \ f(x)\
\tilde{h}(\alpha)\ f(x)^{-1}\ \ .$$ Without loss of generality we
take $f(0) = 1\mkern-6mu 1$.

Now consider the continuous map
$ {h'}_o \ (x)= f(x)\ h_o\ ,$ where $f(x)$ acts on $h_o$ by left
multiplication. This interpolates between the Alice string's Wilson
loop $h_o$, and group element $ {h'}_o (1)$ in the same disconnected
component of H.  $ {h'}_o  (1)$ thus defines a topologically
equivalent string, with Wilson loop $U(2\pi) ={h'}_o  (1)$. In circumnavigating this deformed string, the
original monopole $h(\alpha)$ is unchanged: it goes to
\begin{eqnarray*}\tilde{h'} (\alpha) &=& {h'}_o \, (1)\ h(\alpha)\ {h'}_o^{-1}  \, (1)= f(1)\ h_o\
h(\alpha)\ {h}_o^{-1}\ f(1)^{-1}\\ &=& f(1)\ \tilde{h}(\alpha)\
f(1)^{-1} = h(\alpha)\ \ , \end{eqnarray*}
by construction of $f(x)$. Thus the
monopole loop $h(\alpha)$ remains identical on circumnavigating the
string. Choosing as our nontrivial loop $h(\alpha) = e^{i\alpha
T_h}$, the loop (at each value of $\alpha$) remains unchanged only if
the generator $T_h$ remains unchanged. Thus by continuously deforming
our Alice string's flux from $h_o$ to ${h'}_o  (1)$, we have obtained a
string flux $U(2\pi) = {h'}_o  (1)$ which commutes with all generators;
that is, we have removed all Alice behavior of the string. This renders Alice
behavior for  strings failing our criterion  topologically
unstable; it can be deformed away, and stabilized only in
dynamical, model-dependent ways.

\section{\label{monopoles}Twisted Alice loops as monopoles}

Monopoles lie on the vacuum manifold at spatial infinity, with
topology given by $\pi_2(G/H)$. We here show that a fully twisted
topologically Alice loop is necessarily a topological monopole; that is,
an infinite sphere enclosing it has nontrivial $\pi_2(G/H)$.

First, we construct a sensible twisted Alice loop.

Recall that our Alice string has a condensate $\vev$ which winds
asymptotically over the vacuum manifold $G/H$ according to
$$ \vevphi = U(\varphi)\ \vev_o\ ,
$$
where the Wilson line $U(\varphi)$  acts on the vev $\vev_o$ according to its group representation.  $U(\varphi)$
 varies  continuously over $G$ for $0<\varphi<2\pi$, and connects the identity at $\varphi =0$ to a distinct Wilson loop $U(2\pi)$ in $H$. The string
is topological when $U(2\pi) = h_o$ lies in a disconnected component of $H$,
with nontrivial $\pi_o(H)$, and  is topologically Alice when it meets
our criterion (\ref{crit}).

Now twist the Alice string: continuously rotate its Wilson line within $G$ by  the angle-dependent $H$-group rotation $h^{-1}(\alpha)$:
$$U(\varphi, \alpha) = h^{-1}(\alpha)\ U(\varphi)\ h(\alpha)\ \ ,$$
as shown in Figure 1a. This, of course, rotates our condensate among the degenerate vacua on $G/H$:
$$ \vevphial = U(\varphi, \alpha)\ \vev_o\ \ .
$$

Under what conditions may we identify string ends at $\alpha = 0$ and
$\alpha = 2\pi$ to form a string loop, as pictured in Figure 1b?
First, we require the string configurations to match at the
junction. This is assured if $h(2\pi) = h(0)$, that is, if $h(\alpha)$
is a loop. Second, the twisted condensate $\vevphial$ must be single-valued. Note that the Wilson line $U(\varphi, \alpha)$ itself need not be single-valued: indeed, for a monopole configuration, $U(\varphi, \alpha)$ interpolates from the identity at $\varphi =  0$ to a nontrivial loop in $H$ at $\varphi = 2\pi$.

First we check singlevaluedness of $\vevphial$ at the loop's origin. Here $\varphi = 0$ (or $2\pi$) while
$\alpha$ is indeterminate. Note that $ U(0, \alpha)$ is the identity, manifestly single-valued. At $\varphi = 2\pi,$
$$U(2\pi, \alpha) =  h^{-1}(\alpha)\ U(2\pi)\ h(\alpha)\ \ .$$
This  generally does vary with $\alpha$; however,  it is a loop in $H$, with basepoint $U(2\pi) = h_o \in H$. It thus leaves the condensate invariant, assuming the single value $\vev_o$  at loop origin.

Elsewhere, we need only show first, that $\langle\ \phi\ (\varphi,
\alpha + 2\pi)\ \rangle = \vevphial$; and second, that $\langle\ \phi\
(\varphi + 2\pi, \alpha)\ \rangle = \vevphial$. The first is trivial:
since $h(\alpha)$ is a loop, $h(\alpha) = h(\alpha + 2\pi)$ and both
$U(\varphi, \alpha)$ and $\vevphial$ are single-valued in $\alpha$.

To show singlevaluedness in $\varphi$, let us, without loss of generality, diagonalize our string Wilson line $U(\varphi)$, taking it to be generated by a fixed generator so that
$U(\varphi +  2\pi ) = U(\varphi)\ U(2\pi)\ .$
Then our twisted Wilson line obeys
$$U(\varphi +  2\pi, \alpha) =  h^{-1}(\alpha)\,\, U(\varphi)\,\, U(2\pi)\,\, h(\alpha)
= U(\varphi, \alpha)\,\, U( 2\pi, \alpha)\  .$$
As noted above, $U( 2\pi, \alpha)$ is a loop in $H$, leaving $\vev_o$ invariant. Thus
$$\langle\ \phi\ (\varphi, \alpha + 2\pi)\ \rangle = \vevphial =  U(\varphi, \alpha)\ \vev_o\ \ $$
and our twisted Alice loop is fully single-valued.

By the exact sequence for $\pi_2 (G/H)$, our twisted Alice loop is a
monopole when $U(\varphi, \alpha)$ interpolates between an element of $H$ at
$\varphi = 0$ and a nontrivial loop in $H$ at $\varphi = 2\pi$. For convenience, right multiply  $U(\varphi, \alpha)$ by $h_o^{-1}$:
\beq U(\varphi, \alpha) = h^{-1}(\alpha)\ U(\varphi)\ h(\alpha)\ h_o^{-1}\ \
.\label{twistmon}\eeq
Since $h_o^{-1} \in H$, this right multiplication does not change the physical condensate $\vevphial$. However, it makes the topology of $U(\varphi,\alpha)$  clear, for
$$U(\varphi,\alpha) = \left\{\ \begin{array}{ll} h_o^{-1}& \mbox{for}\ \varphi = 0\\ h^{-1}(\alpha)\ \tilde{h}(\alpha)& \mbox{for}\ \varphi = 2\pi\end{array}\right. \ \ .$$
By definition, if the string is topologically Alice,
$\tilde{h} (\alpha) \ \not\sim \
h(\alpha)$ so that $ h^{-1}(\alpha)\ \tilde{h}(\alpha)$ is a nontrivial loop in $H$, and  the twisted Alice loop carries nontrivial monopole charge. If the string is {\em not} topologically Alice, the loop $ h^{-1}(\alpha)\ \tilde{h}(\alpha)$ is trivial in $H$ and the twisted Alice loop carries no monopole charge.

Thus twisted Alice loops carry monopole charge if and only if they
obey our topological Alice criterion. We note that the
monopole charge displayed , with winding $ h^{-1}(\alpha)\ \tilde{h}(\alpha)$, is
exactly that deposited on an initially untwisted Alice loop,
when a monopole of winding $h^{-1}(\alpha)$ circumnavigates the string
and emerges with winding $\tilde{h}^{-1}(\alpha)$. \cite{blp} (The {\em inverse} twisted
Alice loop, generated by $U^{-1}\ (\varphi, \alpha)$, instead carries
monopole charge $h(\alpha) \
\tilde{h}^{-1}(\alpha)$, deposited in the monopole circumnavigation
$h(\alpha) \ \rightarrow \ \tilde{h}(\alpha)$.)

We note that our final map $U(2\pi,\alpha)$ for the twisted
Alice loop coincides with the flux loop (4.1) and paths $C'_\alpha$
defined in \cite{blp}. They show that this map coincides with the
Lubkin classification of monopole charge for the twisted Alice loop.
\comment{which considers only flux variation over a 1-parameter loop of closed
paths on the sphere at spatial infinity, enclosing the Alice
loop.} This reinforces our classification, as  identifying
twisted topologically Alice loops with physical gauged magnetic monopoles.

We note two key points. First, the topological loop twisting
(\ref{twistmon}) supports only {\bf deposited} monopole charge:
namely, that deposited when a monopole of winding $h^{-1}(\alpha)$
scatters into one of distinct winding
$\tilde{h}^{-1}(\alpha)$. Typical Alice string scattering changes
monopole to antimonopole, with monopole number deposited onto the
string loop in units of two. Our topological arguments thus leave open
the question of whether twisted Alice loops can carry {\bf
fundamental} monopole charge. Second, we implicitly took as twisting function
$h(\alpha)$ in equation (\ref{twistmon}), the loop in $H$ representing
a fundamental monopole.  (We call this choice the fully-twisted Alice
loop). We remain free to choose a different twisting function
$h(\alpha)$ in $H$, so long as it renders $U(\varphi, \alpha)$
singlevalued in the angle $\alpha$. Propitious choice of $h(\alpha)$
in the condensed matter models below allows construction of twisted Alice
loops with {\bf fundamental} monopole charge.

\section{\label{cm}Condensed Matter Alice Strings}

\subsection{\label{schwarz}The Schwarz, or Nematic,  Alice String}

We start with the simplest example, the canonical Schwarz Alice
string, \cite{oldAlice} whose symmetry-breaking pattern coincides with
Alice strings in nematic liquid crystals and in non-chiral Bose
condensates. \cite{leonvol}

Here $G$ is $SO(3)$, with
Higgs $\phi$ transforming in the adjoint.
$\phi$ develops the vev
$\vev = {\rm diag}\ (1,1,-2)\ ,
$
breaking $SO(3)$ to the residual symmetry $H = O(2)$, containing  z-rotations $R_z\,(\alpha)$ and the discrete symmetry element
$ h_o = R_x\, (\pi) = {\rm diag}\ (1, -1, -1)\ .$
Here $\pi_o (H) = Z_2$ and $\pi_1(H) = Z$ so we have topological strings and monopoles. The Alice string has Wilson line
$ U(\varphi) = R_x\, (\varphi/2)$
with $U(2\pi) = h_o$. $U(2\pi)$ fails to commute
with unbroken symmetry generator $T_z$; in fact, on parallel
transport around the string,
\beq T_z \rightarrow U\ (2\pi) \ \ T_z \ \ U^{-1}\ (2\pi) = -T_z \ \ .
\label{canAliceT}\eeq

This Schwarz Alice string meets our topological criterion, of changing
topological monopole charge on circumnavigation. By the exact
sequence for $\pi_2 (G/H)$, topological monopoles are associated with
nontrivial loops in $O(2)$ which can be unwound in $SO(3)$. Since only
even winding loops in $O(2)$ can be unwound in $SO(3)$, the
fundamental monopole in this canonical Alice model has a loop in
$O(2)$ of winding 2.

To apply our topological criterion, we
represent the string by $h_o$, a nontrivial element of
$\pi_o (H)$, and the fundamental monopole by $h(\alpha) = R_z(2\alpha)$, a winding 2 element of $\pi_1(H)$. This gives
$$\tilde{h} (\alpha) = h_o\ h(\alpha)\ h_o^{-1}  = h^{-1} \, (\alpha)\ \ ,$$
from equation (\ref{canAliceT}). Note that $h^{-1} \, (\alpha)$ has $O(2)$ winding -2, topologically distinct from $h(\alpha)$ of $O(2)$ winding 2. Thus
$ \tilde{h} (\alpha)\ \not\sim \
h(\alpha)$ and our topological criterion is met.

We now construct a monopole as a twisted Alice loop. From Eq.~(\ref{twistmon}), the twisted Wilson line
$$ U(\varphi, \alpha) = h^{-1}(\alpha/2)\ U(\varphi)\ h(\alpha/2)\ h_o^{-1}\ \
$$
gives an Alice loop with single-valued condensate. (We take
$h(\alpha/2)$ because $h$ need only  be single-valued in
$\alpha$, and $h(\alpha/2)$, the winding 1 loop in $O(2)$, first
achieves this.)  $U(\varphi, \alpha)$ interpolates between
$ h_o^{-1}$ at $\varphi = 0$ and $ h^{-1}(\alpha)$ at $\varphi =
2\pi$. It is thus the fundamental antimonopole in the model,
winding $-2$ in $O(2)$. The inverse twisted Alice loop, with Wilson
line $ U^{-1}(\varphi, \alpha)$, creates the fundamental monopole.

As we show in the next section, twisted Alice loops in $^3$He-A and
amorphous chiral superconductors also support fundamental monopole
charge. Again, this stems from the fact that monopoles must carry even
$U(1)$ winding, as only even $U(1)$ loops can unwind inside the
$SO(3)$ factors endemic to $G$ for condensed matter
systems. Single-valued half-twisted loops can thus arise, supporting
fundamental monopole charge.  Embedding $H = O(2)$ in a different $G$,
however, can result in Alice loops unable to support fundamental
monopole charge, as we see in section \ref{toponly}.

\subsection{\label{he3a}The  Alice String of $^3$He-A}

A more complicated global symmetry-breaking pattern describes the
Alice string expected in $^3$He-A \cite{volmin,volbook}, and more recently
predicted in amorphous chiral superconductors with p-wave pairing,
such as Sr$_2$RuO$_4$.\cite{spintrips}

Here $G$ is $SO(3)_L\times SO(3)_S\times U(1)_N$, describing spatial
rotations, spin rotations, and a $U(1)$ phase symmetry associated with
number conservation of helium atoms. (The U(1) symmetry is
approximate, as is independence of spin and orbital rotations due to
minimal spin-orbit coupling, but both describe $^3$He-A well.)  The
matrix order parameter $A$ transforms under symmetry transformations
as
$A \ \rightarrow e^{2i\theta}\ R_S \ A\ R_L^{-1}\ ,$
where $R_S$ and $R_L$ are spin and orbital rotations, respectively.

The order parameter develops the form
$$A_{ij} = \Delta_A\ \hat{d}_i\ (\hat{m}_j + i\,\hat{n}_j)\ \ ,$$ where
$\hat{m}$ and $\hat{n}$ are perpendicular, determining $\hat{l} =
\hat{m} \times\hat{n},$ the direction of the
condensate's angular momentum vector. This breaks $G$ to the residual symmetry
$H= U(1)_{S_{\hat{d}}} \ \times\ U(1)_{L_{\hat{l}} - N/2}\ \times Z_2,$
consisting of spin rotations about the $\hat{d}$ axis; spatial
rotations about the $\hat{l}$ axis when compensated by a matching
$U(1)_N$ phase rotation, and the discrete $Z_2$ transformation $h_o$, with
$h_o: \hat{d}, \ \ \hat{m} + i\,\hat{n}\ \rightarrow \   - \hat{d}, \ \
 -\ (\hat{m} + i\,\hat{n}) \ .$

Identifying the defect topology requires care in this setting, as the
exact sequences relating $\pi_2(G/H)$ and $\pi_1(G/H)$, the monopole
and string homotopy groups, to $\pi_1(H)$ and $\pi_0(H)$ are highly
nontrivial.  Note that $\pi_2(G/H) = Z$, corresponding to the loops
$\pi_1( U(1)_{S_{\hat{d}}} )$, which can be unwound in $G$. (Loops of
the other $U(1)$ factor cannot be unwound in $G$, as they contain
unshrinkable $U(1)_N$ loops.) $\pi_1(G/H) = Z_4$, which describes
strings of two different origins. First, the Alice strings, called
half-quantum vortices, have Wilson lines ending in a disconnected
component of $H$, getting topological stability from
$\pi_0(H)$. Second, a $Z_2$ winding one vortex, nontrivial in
$SO(3)_L$ in $G$, induces as its image a $Z_2$ winding one vortex in
$G/H$, with topological stability inherited from $\pi_1 (G)$. These
two classes of vortices are not independent: instead winding twice
about a half-quantum vortex is equivalent to once around a winding
one vortex, and the full string homotopy is $\pi_1(G/H) = Z_4$, or
windings $0, \pm 1/2$, and $1$ modulo 2, with Alice strings corresponding to
windings $\pm 1/2$.

The Volovik-Mineev Alice string, of winding $\pm 1/2$, has order
parameter $A_{ij}$ with $\hat{d} = \hat{x}$ in spin space, and
$\{\hat{l}, \hat{m},\hat{n}\} = \{\hat{x}, \hat{y}, \hat{z}\}$ in ordinary
space. This is acted on by Wilson line
$ U(\varphi) =  e^{\pm i\varphi/2}\ R_{S_{\hat{z}}}(\varphi/2)$
to give, asymptotically in $r$,
$$A_{ij} (\varphi) = \Delta_A\  e^{\pm i\varphi/2}\ (\cos (\varphi/2)\ \hat{x}_j + \sin  (\varphi/2)\ \hat{y}_j)_S\ (\hat{x}_j + i\hat{y}_j)_L\ \ ,$$
single-valued in $\varphi$. Note that  $U(2\pi) = - R_{S_{\hat{z}}}(\pi) $ lies in the same homotopy class as $h_o$. This string is Alice, making
unbroken symmetry generator $T_{S_{\hat{x}}}$ double-valued.
Physically, this means that a particle flips its spin, and hence its magnetization, on circumnavigating the Alice string.

This long-studied Alice string meets our topological criterion, of changing
topological monopole charge upon circumnavigation. By the exact
sequence for $\pi_2 (G/H)$, topological monopoles are associated with
nontrivial loops in $ U(1)_{S_{\hat{x}}}$ which can be unwound in $SO(3)_S$. As in the nematic case,  only
even winding loops in $ U(1)_{S_{\hat{x}}}$ can be unwound in $SO(3)_S$. Thus the
fundamental monopole in $^3$He-A corresponds to  a loop in
$ U(1)_{S_{\hat{x}}}$ of winding 2.

In applying our topological criterion, we choose $U(2\pi)$ as our
representative of the string, a nontrivial element of
$\pi_o (H)$, and $h(\alpha) = R_{S_{\hat{x}}}(2\alpha)$ as our representative of
the fundamental monopole, a winding 2 element of $\pi_1( U(1)_{S_{\hat{x}}})$. This gives
$$\tilde{h} (\alpha) = h_o\ h(\alpha)\ h_o^{-1}  = h^{-1} \, (\alpha)\ \ ,$$
since $T_{S_{\hat{x}}}\rightarrow - T_{S_{\hat{x}}}$. Note that $h^{-1} \, (\alpha)$ has $U(1)_{S_{\hat{x}}}$ winding -2, topologically distinct from $h(\alpha)$ of $U(1)_{S_{\hat{x}}}$ with winding 2. Thus
$ \tilde{h} (\alpha)\ \not\sim \
h(\alpha)$, meeting our topological criterion.

As in the nematic case, we construct a monopole as a twisted Alice loop. From Eq.~(\ref{twistmon}), the twisted Wilson line
$$ U(\varphi, \alpha) = h^{-1}(\alpha/2)\ U(\varphi)\ h(\alpha/2)\ h_o^{-1}\ \
$$
generates an Alice loop with single-valued condensate. (Again
$h(\alpha/2)$ appears,  the winding 1 loop in $ U(1)_{S_{\hat{x}}}$, as our minimal single-valued choice in  constructing  $U(\varphi, \alpha)$.)  $U(\varphi, \alpha)$ interpolates between
$ h_o^{-1}$ at $\varphi = 0$ and $ h^{-1}(\alpha)$ at $\varphi =
2\pi$. It is thus the fundamental antimonopole in the model, which
contains monopoles and antimonopoles of even winding in $ U(1)_{S_{\hat{x}}}$
only. Note that the inverse twisted Alice loop, with twisted Wilson
line $ U^{-1}(\varphi, \alpha)$, again generates the fundamental monopole.

\section{\label{partmodels} Particle physics  Alice strings}
\subsection{\label{nontop}A Nontopologically Alice string}

Consider the nontopologically Alice string introduced
in \cite{stringzm}: a Higgs $\phi$, transforming
in the adjoint under $G=SO(6)$, acquires the vev $\vev
= {\rm diag}(1^3,-1^3)$. This
condensate leaves unbroken an $SO(3)\times SO(3)$ subgroup of $SO(6)$
and a discrete $Z_2$ transformation $h_1=-1\mkern-6mu 1_6$, so
$H=SO(3)\times SO(3)\times Z_2$.  Here $\pi_o(H)=Z_2$ and $\pi_1(H) =
Z_2 \times Z_2$, so topological strings and monopoles form,
with monopoles and antimonopoles identified.   Alice characterististics of the string
depend on $U(2\pi)$.  For $U(2\pi) = h_1$,
 all unbroken generators $T_{ij}$ are single-valued under parallel transport around the string,
and the string is not Alice. However, for the topologically equivalent choice
$U(2\pi) = {\rm diag} (1^2, (-1)^4) = - R_{12}(\pi) $,
the
string is algebraically Alice, making  generators $T_{13}$ and $T_{23}$ of rotations in
the $13$- and $23$-planes double-valued.
Since this Alice behavior is removable by deforming to the
topologically equivalent string with $U(2\pi) = h_1$, it must be
nontopological. However, it is instructive to see how the two strings
fail our topological criterion for Alice behavior. We take as our  nontrivial monopole loop $h(\alpha) = R_{13}\ (\alpha)$, with monopole charge
(1,0). We find, for the string with $h_o = h_1$,
$\tilde{h} (\alpha) = h_o\ h(\alpha)\ h_o^{-1} = h \, (\alpha).$
That is, our topological criterion fails, as the monopole remains
unchanged in circumnavigating the Alice string. 
Instead, for the Alice string with 
$h_o =  - R_{12}(\pi)$, 
$\ \tilde{h} (\alpha) = h_o\ h(\alpha)\ h_o^{-1} = h^{-1} \, (\alpha)\
,$ since $T_{13}\rightarrow -\ T_{13}$. Here the (1,0) monopole transforms into its
antimonopole on traversing the string. {\bf However,} that
transformation is nontopological, as monopoles and antimonopoles are
topologically equivalent. So, despite algebraic Alice
behavior, this string also is not topologically Alice. Loops
$h(\alpha)$ alter on string traversals, but in a
topologically trivial way.

We might still hope to construct a (1,0) monopole as a twisted string
loop, taking for our string the algebraic, but nontopologically Alice
string $U(\varphi) = R_{34}(\varphi/2) \ R_{56}(\varphi/2)$, with
algebraic Alice flux $U(2\pi) = h_o = - R_{12}(\pi) $ as above. From
Eq.~(\ref{twistmon}),the twisted Wilson line
$$ U(\varphi, \alpha) = h^{-1}(\alpha)\ U(\varphi)\ h(\alpha)\ h_o^{-1}\ \
$$
generates an Alice loop with single-valued condensate.  $U(\varphi,
\alpha)$ interpolates between $ h_o^{-1}$ at $\varphi = 0$ and $
h^{-2}(\alpha)$ at $\varphi = 2\pi$. This is a loop in $H$ of winding
(-2,0); however, winding (-2,0) loops are deformable to the identity in $H$, so
this fully twisted nontopologically Alice loop fails to carry topological
monopole charge.

Recall that, in building singlevalued twisted Alice loops, we required
$ U(\varphi, \alpha)$ to be singlevalued in $\alpha$; we thus
identified $h(\alpha)$ as a loop in $H$. Strictly, we do not need
$h(\alpha)$ to be a loop; all we need is
\beq h^{-1}(2\pi)\ U(\varphi)\ h(2\pi) = U(\varphi)\  .\label{singval}\eeq
We might still hope to build the  fundamental (1,0) monopole as a twisted Alice loop, exploiting this freedom in $h(\alpha)$. Were the twisted loop
$$ U(\varphi, \alpha) = h^{-1}(\alpha/2)\ U(\varphi)\ h(\alpha/2)\ h_o^{-1}\ \
$$
single-valued, it would carry fundamental (-1,0) monopole charge, as it  interpolates between
$ h_o^{-1}$ at $\varphi = 0$ and the nontrivial (-1,0) antimonopole $ h^{-1}(\alpha)$ at $\varphi =
2\pi$. However, this twisted loop candidate is not single-valued; it obeys instead
$ h^{-1}(2\pi)\ U(\varphi)\ h(2\pi) = U^{-1}(\varphi)\ .$ We
thus cannot build a fundamental (1,0) monopole as a twisted Alice loop in
this model, where Alice behavior is nontopological and monopole
charge is $Z_2 \times Z_2$.

This possibility to construct $U(\varphi,\alpha)$ single-valued in
$\alpha$, {\bf without} forcing $h(\alpha)$ to be a loop, always
merits investigating. Indeed, in \cite{skyrme, global}, one of us
constructed what is essentially the fundamental monopole in this
model, by exploiting exactly such an accidental algebraic
singlevaluedness. That construction (most clearly in section IIIA of
\cite{global}, taking $F(r), \varphi$ as the spherical coordinates
$\theta, \varphi$ at spatial infinity),is quite similar to the
twisting constructions here. However, it describes a fundamentally
point-like defect, and cannot be interpreted as a twisted loop.

\subsection{\label{toponly}A Topologically Alice Loop Carrying only Deposited Monopole Charge }

We consider a slightly modified canonical Alice string. Take $G$ to be
$SU(3)$, with Higgs $\phi$ transforming according to $\phi \rightarrow
g\, \phi\, g^T$ (giving fermions in this model a Majorana mass).
When $\phi$ develops the vev
$\vev = {\rm diag}\ (1,1,-2)\ \ ,
$
$SU(3)$ breaks to the residual symmetry $H = O(2)$,
identical to that of the canonical Schwarz Alice string.
Again we have $\pi_o (H) = Z_2$ and $\pi_1(H) = Z$,
with topological strings and monopoles. We have the same Alice string
as in the canonical case, making the $O(2)$ generator $T_z$
double-valued. This Alice behavior is again topological, as our topological
criterion, that $\pi_o(H)$ acts nontrivially on $\pi_1(H)$, depends
only on the unbroken symmetry group $H$.

Where we deviate from the canonical Alice string model is in the
identification of twisted Alice loops as monopoles. Here, by the exact
sequence for $\pi_2 (G/H)$, topological monopoles are associated with
nontrivial loops in $O(2)$ which can be unwound in $G$, here
$SU(3)$. {\bf All} nontrivial loops in $O(2)$ can be unwound in
$SU(3)$; thus the fundamental monopole in this model  has a loop
in $O(2)$ of winding 1.

We now construct a monopole as a twisted Alice loop. We take
$U(\varphi) = R_x\, (\varphi/2)$, as in the canonical Alice case, and
$h(\alpha) = R_z(\alpha)$, a loop of winding 1. From
Eq.~(\ref{twistmon}), the twisted Wilson line
$$ U(\varphi, \alpha) = h^{-1}(\alpha)\ U(\varphi)\ h(\alpha)\ h_o^{-1}\ \
$$ generates a twisted Alice loop with single-valued condensate.
$U(\varphi, \alpha)$ interpolates between $ h_o^{-1}$ at $\varphi = 0$
and $ h^{-2}(\alpha)$ at $\varphi = 2\pi$. This twisted Alice loop
carries monopole charge of $-2$, which while nontrivial is {\bf not}
the fundamental antimonopole in this model. (Similarly, the inverse
twisted Alice loop, with Wilson line $ U^{-1}(\varphi, \alpha)$,
carries monopole charge +2).

Again, we might still hope to build a fundamental monopole as a
twisted Alice loop, by allowing $h(\alpha)$ above to be not a loop,
but a curve obeying Eq.~(\ref{singval})
This looser
constraint still guarantees singlevaluedness in $\alpha$ of $
U(\varphi, \alpha)$.  Indeed, were the half-twisted loop
$$ U(\varphi, \alpha) = h^{-1}(\alpha/2)\ U(\varphi)\ h(\alpha/2)\ h_o^{-1}\ \
$$ single-valued in $\alpha$, with $h(\alpha) = R_z(\alpha)$ as above,
it would carry fundamental antimonopole charge. This is because it
interpolates between $ h_o^{-1}$ at $\varphi = 0$ and the winding $-1$
loop $ h^{-1}(\alpha)$ at $\varphi = 2\pi$. However, this half-twisted loop
candidate is not single-valued in $\alpha$; it obeys instead
$ U(\varphi, 2\pi) = R_x^{-1} (\varphi) \ U(\varphi, 0)\  .
$ We thus cannot build a fundamental monopole as a twisted Alice loop
in this model. Instead twisted Alice loops carry only the monopole
charge which topological arguments ensure they must carry: because
monopoles scatter into antimonopoles on transiting Alice loops, Alice
loops {\bf must} support deposited monopole charge, which arises in
units of 2. Our full twisting construction creates twisted Alice loops
supporting exactly that deposited charge.

\section{\label{conclusions}Conclusions}

We have established a topological criterion for strings to display
Alice behavior. This criterion, that $\pi_o(H)$ acts nontrivially on
$\pi_1(H)$, depends only on the residual symmetry group $H$. Alice
strings must form in models obeying this criterion, while Alice
behavior can be deformed away for strings failing the
criterion. Particularly, the criterion requires that topological
monopoles always accompany topologically Alice strings; and
furthermore, that topologically Alice strings alter the topological
charge of monopoles that circumnavigate them. We construct monopoles
as twisted loops of Alice string, and show that such twisted loops can
always support deposited monopole charge. Whether twisted Alice loops
can support fundamental monopole charge depends on the
symmetry-breaking pattern more closely, as we examined in condensed
matter and illustrative particle physics models.
Specifically, it depends on the initial
symmetry group $G$, through the identification of loops in $H$ with
monopoles via the exact sequence for $\pi_2(G/H)$. When fundamental
monopoles correspond to nonminimal-winding loops in $H$, 
single-valued half-twisted Alice loops may arise, carrying fundamental monopole
charge. This occurs generically for condensed matter topologically
Alice strings, since only winding 2 loops in $H$ unwind in the $SO(3)$
initial symmetry groups associated with angular momenta.  When
fundamental monopoles, instead, have minimal winding in $H$, only
algebraic accident allows a half-twisted Alice loop to be
single-valued. Generically, in such minimal embeddings, twisted Alice
loops can support {\bf only} deposited, not fundamental, monopole
charge.

\begin{acknowledgments}
Early stages of this work were supported by NSF grant PHY-9631182 and
by the University Research Committee of Emory University. KB thanks
the KITP (under NSF grant PHY99-07949)  for hospitality during the writing of
an early version of this paper.  The work of T.I. was supported in part by the U.S. Department of Energy.
\end{acknowledgments}

\newpage

%\vspace{4pt}
$$\begin{array}{c}
\epsfig{file=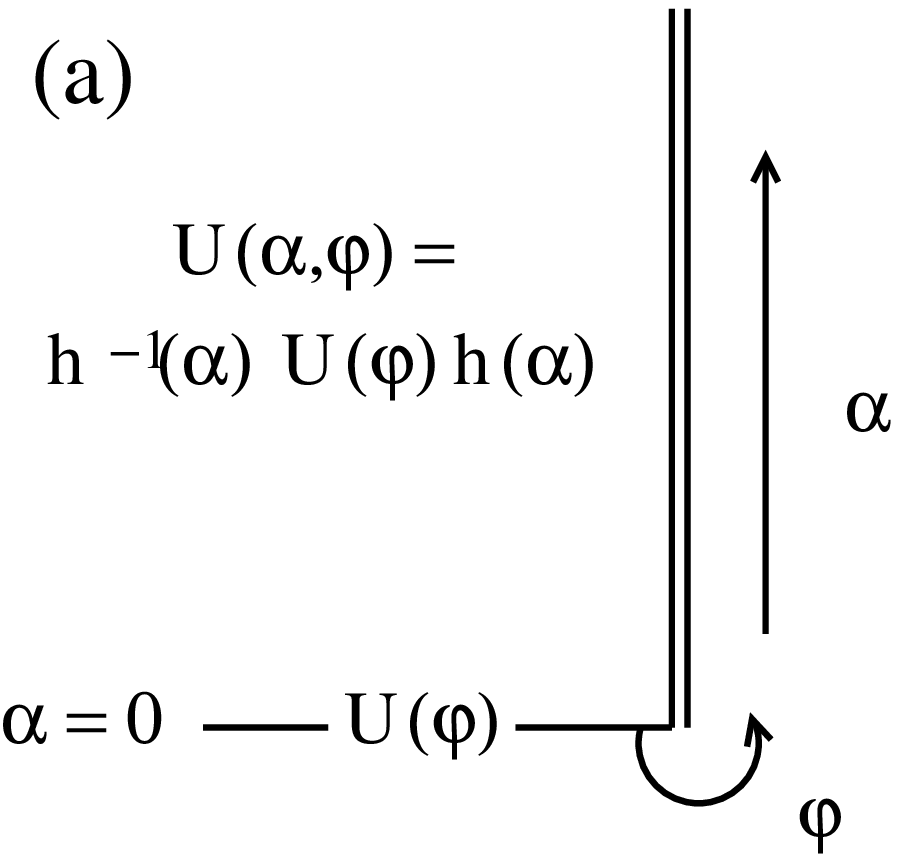, width = 4.0cm}\quad\quad\epsfig{file=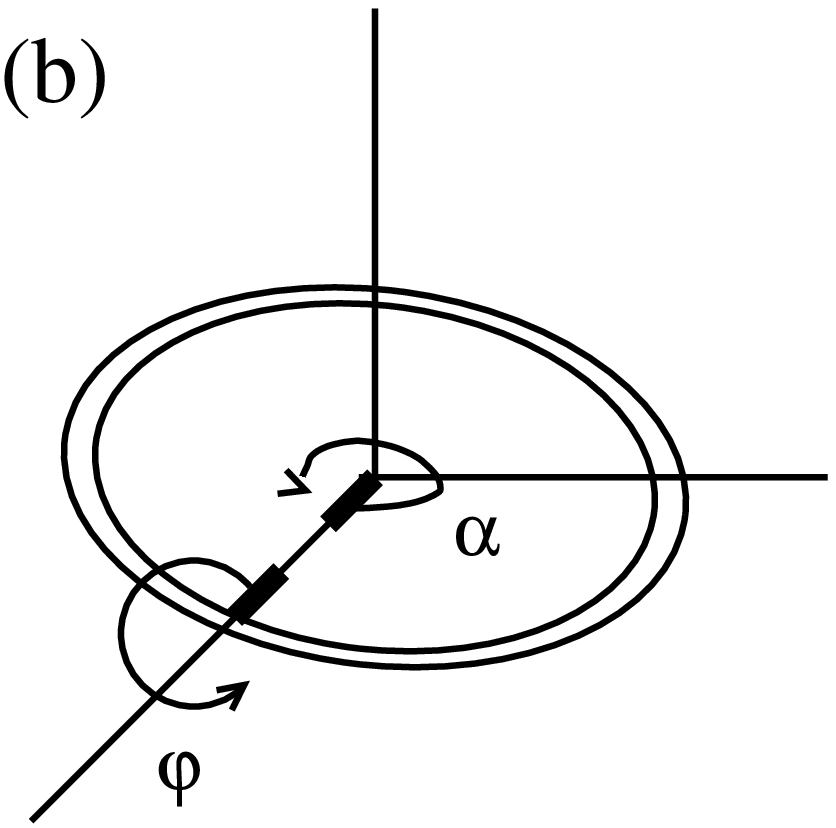, width = 4.0cm}\\
\parbox{8.6cm}{FIG. 1: a) A twisted Alice string. b) Identifying twisted Alice string ends to form a twisted Alice loop.}
\end{array}$$
%\vspace{4pt}


\newpage

\begin{thebibliography}{9999}
\baselineskip 16pt
%
%
\bibitem{oldAlice} A.~S.~Schwarz,
{\em Nucl.\ Phys.\ }{\bf B208} (1982) 141;
A.~Balachandran, F.~Lizzi, V.~Rodgers,
{\em Phys.\ Rev.\ Lett.\ }{\bf 52}
(1984) 1818.
%
%
\bibitem{stringzm} M. Alford, K. Benson, S. Coleman, J. March-Russell
and F. Wilczek,
{\em Phys. Rev. Lett.\ } {\bf 64} (1990) 1632;
 {\em
Nucl. Phys.\ }{\bf B349} (1991) 414.
%
\bibitem{newAlice} J.~Preskill and L.~Krauss,
{\em Nucl.\
Phys.\ }{\bf B341} (1990) 50.
%
%
\bibitem{blp}  M.~Bucher, H.~K.~Lo, and J.~Preskill,
{\em Nucl.\ Phys.\ }{\bf B386} (1992) 3.
%
%\cite{ABrefs}
\bibitem{ABrefs}
P.~McGraw,
{\em Phys.\ Rev.\ }{\bf D50} (1994) 952;
 J. March-Russell and F. Wilczek,  {\em
Phys. Rev. Lett.\ } {\bf 68} (1992) 2567; A.~Davis and A.~Martin, {\em Nucl.\ Phys.\ }{\bf B419} (1994) 341.
%
%\cite{global}
\bibitem{global}
K.~Benson,
%``Charge violation and Alice behavior in global and textured strings,''
Phys.\ Rev.\ D {\bf 64}, 085002 (2001)
[arXiv:hep-th/0103238].
%%CITATION = HEP-TH 0103238;%%
%
%
%\cite{volbook}
\bibitem{volbook}
G.~Volovik,
{\bf The Universe in a Helium Droplet} Oxford University Press (2003)
[http://ice.hut.fi/Volovik/book.pdf].
%
%
%\cite{leonvol}
\bibitem{leonvol}
U.~Leonhardt and G.~E.~Volovik,
%``How to create Alice string (half-quantum vortex) in a vector  Bose-Einstein condensate,''
JETP Lett.\  {\bf 72}, 46 (2000)
[arXiv:cond-mat/0003428].
%%CITATION = COND-MAT 0003428;%%
%
\bibitem{volmin}
G.~Volovik and V.~Mineev, JETP {\it Lett.}
{\bf 24}, 561 (1976).
%%CITATION = ZFPRA,24,561;%%
%
\bibitem{spintrips}
H.~Kee, Y.~Kim, and K.~Maki,
%Half-quantum vortex and d-soliton in Sr$_2$RuO$_4$
Phys. Rev. B 62, R9275 (2000)
[arXiv:cond-mat/0005510];
T. M. Rive and M. Sigrist, J. Phys. Cond. Matter, 7, L643 (1995);
M. Sigrist et al, Physica C 317-318, 134 (1999).
%
%
%
\bibitem{semew}
A.~Achucarro and T.~Vachaspati,
%``Semilocal and electroweak strings,''
Phys.\ Rept.\  {\bf 327}, 347 (2000)
[arXiv:hep-ph/9904229].
%
%
\bibitem{skyrme} K. Benson, A. Manohar, and M. Saadi,  {\em Phys. Rev. Lett.\ } {\bf 74} (1995)
 1932 [arXiv:hep-th/9409042]; K. Benson and M. Saadi,
  {\em Phys.
 Rev.\ }{\bf D51} (1995) 3096 [arXiv: hep-th/9409109].
%



\end{thebibliography}
\end{document}